\documentclass{aa}  

\usepackage{graphicx}
\usepackage{gensymb}
\usepackage{txfonts}
\usepackage{booktabs}       

\usepackage{adjustbox}
\usepackage{longtable}
\usepackage{supertabular}
\usepackage{caption}
\usepackage[final]{changes}
\usepackage{hyperref}
\hypersetup{
  colorlinks   = true, %Colours links instead of ugly boxes
  urlcolor     = blue, %Colour for external hyperlinks
  linkcolor    = blue, %Colour of internal links
  citecolor   = blue %Colour of citations, could be ``red''
}
\begin{document}

   \title{A systematic search for double eclipsing binaries in Zwicky Transient Facility data}
   
   % \added{\thanks{Table B.1 is only available in electronic form at the CDS via anonymous ftp to cdsarc.u-strasbg.fr (130.79.128.5) or via http://cdsweb.u-strasbg.fr/cgi-bin/qcat?J/A+A/}}}

   \author{T. Vaessen \inst{1}\thanks{t.vaessen90@gmail.com} 
          \and
          J. van Roestel\inst{1}\thanks{j.c.j.vanroestel@uva.nl}
          }

   \institute{Anton Pannekoek Institute for Astronomy, University of Amsterdam, 1090 GE Amsterdam, The Netherlands\
             }

   \date{Received 31 October, 2023; accepted 11 December, 2023}

% \abstract{}{}{}{}{} 
% 5 {} token are mandatory
  
  \abstract
   {Double eclipsing binaries are gravitationally bound quadruple systems in a `2+2' configuration where both of the binaries are eclipsing. These systems are interesting objects to better understand stellar formation, to investigate the dynamical interaction between the two binary systems or to study certain stages of stellar evolution, such as common-envelope events or Type Ia Supernovae.}
   {With this work, we aim to determine if double eclipsing binaries can be found using ZTF data and what the difficulties are in doing so. Secondly, we aim to significantly increase the number of known double eclipsing systems and determine how this sample differs from samples of double eclipsing binaries found with other telescopes.}
   {We develop a new method to systematically search for double eclipsing binaries in sparsely sampled light curves. For this we use box-least-squares (BLS) to search for the period of the first binary in the system. We then remove that signal from the light curves, and search the residual light curve again with BLS to find the second period. We applied this method to ZTF light curves of 575\,526 eclipsing binaries known in the \textit{Gaia} eclipsing binary catalogue.}
   {We report the discovery of 198 new double eclipsing binary systems. The shortest and longest orbital periods of the newly detected systems are 0.11 days to 323 days respectively.}
   {We successfully implemented a method that systematically searches for double eclipsing binary systems in sparsely sampled data. In total 198 new double eclipsing binary systems have been found in 575\,526 light curves ($\approx$ 0.034$\%$). The ZTF sample typically contains more short period binaries compared to the TESS sample, but is also able to find systems with longer periods than what is currently known. 
   We expect that at least three to four times more quadruples can be found by applying this method to all ZTF stellar light curves, by increasing the number of data points as a result of longer observations, and by implementing an automatic detection mechanism that replaces visual inspection.}

    \titlerunning{Double eclipsing binaries from ZTF}
    \authorrunning{T. Vaessen \& J. van Roestel}
   \maketitle
%
%________________________________________________________________
 
\section{Introduction}

Most stars in the \replaced{Universe}{universe} can be found in gravitationally bound binaries or higher-order hierarchies \citep{Tokovinin_2014}. It is estimated that approximately $4\%$ of all solar-type stars can be found in quadruple systems \citep{2021Univ....7..352T}, either in a 3+1 or 2+2 configuration. A small fraction of 2+2 quadruple systems are double eclipsing binaries, in which two eclipsing binaries also orbit each other \citep{Zasche2019}. If the motion of the inner pair is not strongly perturbed by the outer companions, the motions of the stars are approximated by stable Keplerian orbits and can survive for a long time \citep{Tokovinin_2014}.

Compact double eclipsing binaries are interesting objects of study to better understand stellar formation or to investigate the dynamical interaction between the two binary systems. An important example of such a dynamical interaction is the Kozai-Lidov mechanism where, on long timescales, a periodic exchange between the binaries' eccentricities and inclination can take place \citep{1962AJ.....67..591K,LIDOV1962719}. 
Furthermore, these objects serve as an astrophysical laboratory when studying certain stages of stellar evolution, such as common-envelope events or Type Ia Supernovae \citep{kostov_97_2022}.
Therefore, the identification of more double eclipsing binaries will lead to a larger sample that can be used to study the origin and life cycle of these objects.
Quadruples are intrinsically rare and require a favourable alignment of both binaries with the observer in order to be detected. However, in the case of fortunate alignment, the objects can be identified by periodic eclipses in the light curve. \citet{Zasche2019} systematically searched the OGLE-LMC data by visually inspecting the light curves of eclipsing binary systems and found 72 systems of double eclipsing binaries. More recently, light curves of the Transient Exoplanet Survey Satellite (TESS, \citealt{Ricker2015}) have been used to identify double eclipsing binaries in this way. \citet{zasche_multiply_2022} identified 116 double eclipsing binaries using TESS, and \citet{kostov_97_2022, Kostov_101_2022} detected in total 199 double eclipsing binaries, 18 of which overlapped with \citet{zasche_multiply_2022}. 

In this \replaced{paper}{letter}, we present the search for double eclipsing binaries in Zwicky Transient Facility light curves. In Sect. \ref{sec:data} we present the target preselection and the ZTF data. In Sect. \ref{sec:method} we discuss the method we used to identify double eclipsing binaries. In Sect. \ref{sec:analysis} we present the analysis of the data. The results are presented in Sect. \ref{sec:results} and we discuss them further in Sect. \ref{sec:discussion}. We end the paper with a summary and recommendations for future work in Sect. \ref{sec:conlusion}.

\section{Data}\label{sec:data}
As part of the Zwicky Transient Facility (ZTF), the Palomar 48-inch (P48) telescope has been imaging the sky every night since 2018 \citep{graham_zwicky_2019,bellm_zwicky_2019, dekany_zwicky_2020}. Most of the time, ZTF uses the $g$ and $r$ bands, but a small fraction of the observations are also made in the $i$ band. The exposure times are predominantly 30 seconds for $g$ and $r$ and 60 or 90 seconds for $i$ exposures. The median limiting magnitude, averaged over the lunar cycle, is \mbox{$\approx20.5$} in all three bands. We used the PSF-photometry-based light curves which are automatically generated for all persistent sources detected in the ZTF reference images \citep[for a full description see][]{masci2019} and are publicly available.

In order to avoid searching billions of ZTF light curves, we make a preselection of objects. We process only light curves of objects that have been identified as eclipsing binaries by \textit{Gaia} \citep{2023A&A...674A..16M}. Although this preselection inevitably means that we will miss any system that has not been identified as an eclipsing star by \textit{Gaia} first, the advantage is that it strongly reduces the number of light curves we need to search. 

The \textit{Gaia} eclipsing binary catalogue contains 2\,184\,477 sources distributed over the whole sky and down to magnitude $G\approx20.5$. While ZTF goes slightly deeper than \textit{Gaia}, ZTF only covers the sky from the celestial north pole to a declination of $-30\deg$. This means there are 1\,210\,001 sources in the \textit{Gaia} eclipsing binary catalogue that are also in the ZTF footprint and are fainter than $G=12$ (the approximate ZTF saturation limit). To download ZTF light curves for these objects, we used \textit{ ZTFquery}\footnote{\url{https://github.com/MickaelRigault/ztfquery}} \citep{2018zndo...1345222R}.

\section{Method}\label{sec:method}
Light curves of double eclipsing binaries contain the signal of two eclipsing binaries with different periods. If the light curve is well sampled (e.g. the continuous light curves of TESS), multiple epochs sample each individual eclipse and the individual eclipses are easy to identify. This makes the identification of double eclipsing binaries relatively straightforward \citep{Zasche2019,kostov_97_2022,Zasche_2022_firststudy}. However, if a light curve is sparsely sampled (the typical time between observations is much larger than the duration of an eclipse) individual eclipses are sampled by a few or just a single epoch only and the light curve needs to be phase-folded in order to identify eclipses. Therefore, searching for double eclipsing binaries requires a slightly different approach for sparsely sampled light curves like ZTF. Instead of identifying individual eclipses, we use the fact that both eclipse signals are periodic and use period-finding methods to identify both periods.

\begin{figure}[] 
\centering
\includegraphics[width=\columnwidth]{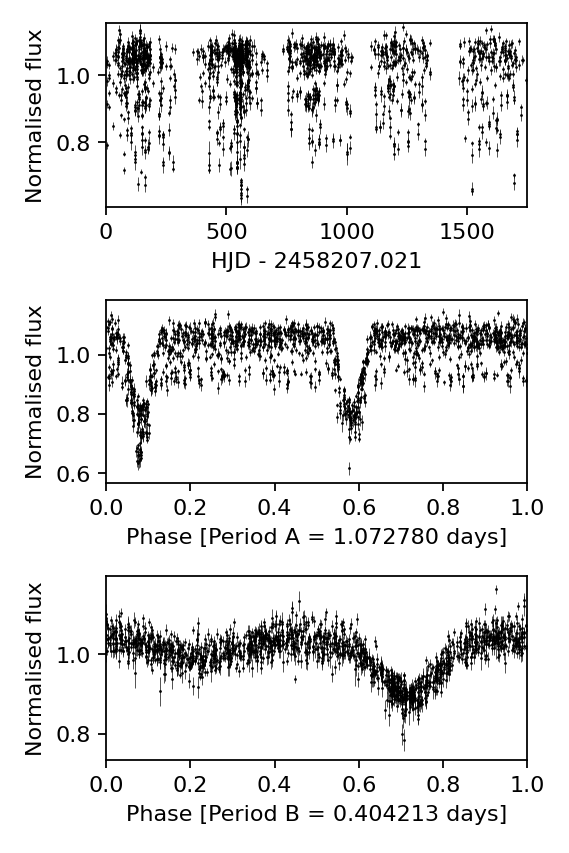}
\caption{ZTF light curves of Gaia dr3 source ID 2216420454386370176. Upper panel: Normalised flux as a function of time. It shows two sets of eclipses with  $\mathrm{P_{A}} = 1.072780$ days and $\mathrm{P_{B}} = 0.404213$ days. Middle panel: Phase folded light curve for binary A with $\mathrm{P_{A}} = 1.072780$ days. We also see that not all data points fall nicely on the curve, indicating the presence of another signal. Lower panel: Phase folded light curve for binary B, after removal of the signal at period A, with $\mathrm{P_{B}} = 0.404213$ days.}
\label{fig: 3 light curves}
\end{figure}

We implemented this method using box-least-squares \citep[BLS,][]{kovacs_box-fitting_2002} which is available through the Python package Astrobase \citep{astrobase}. First the data from the different bands ($g$ and $r$) were combined by normalising the data in each band to its corresponding mean. Then we performed the BLS period search on the light curve to find the first period, which we define as period A. In our experience, this period was often half the orbital period and thus the primary and secondary eclipse overlapped. To avoid this overlap, we phase-folded the light curve to twice period A. To remove the signal of period A, we binned the phase-folded light curve using an empirically determined number of 100 bins. We then divided each flux measurement in the light curve by the mean of its corresponding bin. After the removal of the signal at period A, we applied the BLS method again to find the second period; period B. At each point in this process, we saved the phase-folded and binned light curves for visual inspection. An example of a (phase-folded) light curve can be seen in Fig. \ref{fig: 3 light curves}.

\section{Analysis}\label{sec:analysis}

To ensure that the method worked correctly, we used a test dataset consisting of 68 double eclipsing binaries found in prior research \citep{zasche_multiply_2022} for which ZTF data is available. Furthermore, this test sample also helped to establish a method for identifying double eclipsing binaries quantitatively. For this, the relative peak height (RPH) was introduced as \begin{equation}
\mathrm{RPH} \equiv \frac{\mathrm{peak\ periodpower} - \overline{\mathrm{period power}}}{\sigma_{\mathrm{period power}}}
\end{equation} where max periodpower is the BLS-periodogram peak associated with the best period, $\overline{\mathrm{period power}}$ is the mean of powers and $\sigma_{\mathrm{period power}}$ is the standard deviation. In \textsc{Astrobase} these periodpowers are referred to as lspvals \citep{astrobase}. For readers not familiar with the BLS algorithm, an example of a periodogram is shown in the appendix in Fig. \ref{fig:periodogram}. Here we see the periodogram power value of each orbital period. Additionally, \textsc{Astrobase} offers a function to calculate the signal-to-noise ratio (S/N) for each periodic signal \citep{astrobase}. From the test sample we empirically determined that a RPH > 10 and S/N > 10 may indicate the presence of a \emph{double} eclipsing binary. As it was not feasible to visually inspect all objects in the sample, we selected only candidates with RPH > 10 and S/N > 10. This significantly reduced the number of light curves that needed visual inspection to identify double eclipsing binary candidates. These selection criteria are discussed further in Sect. \ref{sec:discussion}.

Some objects were, based on a high relative peak height and S/N, initially identified as quadruple candidate but appeared to exhibit a sinusoidal rather than an eclipsing signal. These objects were associated with relatively short periods $\mathrm{P_{B}} < 0.30$ days, which could be an indication of a variable star in the binary system that lies on the instability strip, such as $\delta$ Scuti stars \citep{pietrukowicz_over_2020}. 
Another common false positive is associated with the imperfect removal of signal A. This would result in in period B being the same as period A, or an integer multiple of period A. We therefore removed any candidate for which period B was a multiple of period A.

When testing the method, it appeared that light curves containing high cadence observations (nights with more than 30 epochs) had a negative impact on the ability to find the right periods. This is probably caused by correlated noise introduced by the calibration of the data. Therefore, observations with more than 30 epochs per night were removed. A similar effect was observed by outliers in high and low flux. For this reason, outliers in normalised flux were removed using the interquartile range with a cut-off value of 5. Other unreliable data, flagged by the ZTF data process pipeline, was also removed. After this data pre-processing and cleaning, a minimum threshold of 500 data points per light curve was set for analysis. This decreased the sample size from 1\,210\,001 to 575\,526 objects.

Based on the objects in the test sample, we set the period search range for new objects between 0.1 and 150 days with a stepsize determined by the minimum transit duration and observational period. We chose this lower limit as it is not expected to find any objects with shorter orbital periods \citep{drake_ultra-short_2014}. The upper limit was based on the five-year observational period of the ZTF and the decreasing likelihood of finding longer periods. For the transit duration, the BLS performed a search for a periodic signal lasting between 0.01 and 0.2 in phase. 

Finally, we applied the pipeline to the 575\,526 ZTF light curves. We used the HELIOS computer cluster to process the light curves. The average processing time per object is approximately 10 seconds. 

We used the BLS output statistics to selected 5350 promising candidates. These were visually inspected in order to confirm the double eclipsing nature of the objects. At this point, we also determined, to the best of our abilities, what the two orbital periods are, since the BLS algorithm typically finds half the orbital period. Finally, we also checked if the photocenter position was correlated with the eclipse periods to identify chance alignments of two unrelated eclipsing binaries. The typical RMS positional accuracy is $\approx0.1\arcsec$. No chance alignments were found in this manner. 

\section{Results}\label{sec:results}
Of the 575\,526 light curves analysed (500 data points or more), 203 ($\approx$ 0.035$\%$) are identified as double-eclipsing binaries candidates. They are summarised in the appendix in Table \ref{tab:longtable}, and an example is shown in Fig. \ref{fig: 3 light curves}. 
Almost all candidates for double eclipsing binaries are not currently known in any scientific literature \citep[i.e. they are not listed as such on SIMBAD;][]{2000A&AS..143....9W}. Only four of these 203 objects have been found in prior research and marked in Table \ref{tab:longtable} by an asterisk. One of these four objects, \textit{Gaia} dr3 source ID 2003428628134264448, has previously been identified as a triple star system \citep[TIC 388459317;][]{Borkovits_2021} but the ZTF data suggests that it is a double-eclipsing binary. The other three objects, \textit{Gaia} dr3 source ID's 2060949991271681664 \citep{Zasche_2022_firststudy}, 407849617690288128  \citep[TIC 354314226;][]{Kostov_101_2022} and 158123726424621824 \citep[TIC 150055835;][]{Kostov_101_2022} have been identified as double eclipsing binary systems by previous studies that used TESS data. Using the ZTF data, we find that the periods of these objects agree between both studies.

\begin{figure}[h!] 
\centering
\includegraphics[width=\columnwidth]{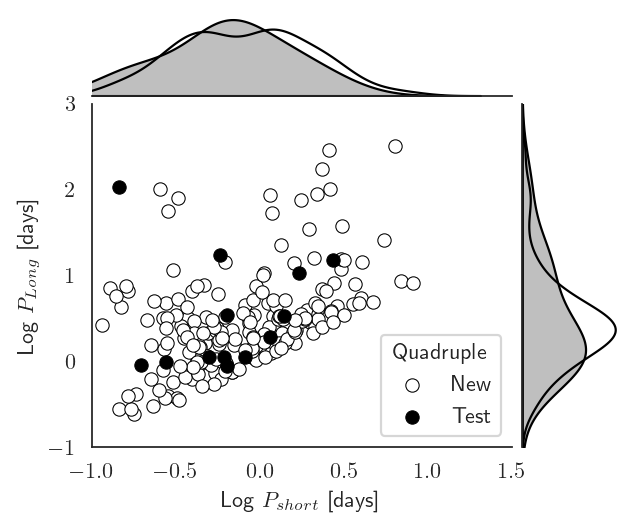}
\caption{Periods A and B of 13 test objects and 203 quadruple candidates. Here $P_{short}$ is defined as the shortest period of A and B and $P_{long}$ as the longest period of A and B. The figure shows that most objects have both periods shorter than a few days. Using a kernel density estimate, this distribution seems to be the same for the candidates and test objects. }
\label{fig: jointplot periods}
\end{figure}

The periods of the double eclipsing binaries in the ZTF sample range from 0.11 days to 323.20 days and the mean of periods A and B are 1.21 days and 10.52 days respectively, see Fig. \ref{fig: jointplot periods}. The figure shows that most objects have both periods shorter than a few days. Using a kernel density estimate, this distribution seems to be the same for the candidates and the test objects.

\begin{figure*}[] 
\includegraphics[width=\textwidth]{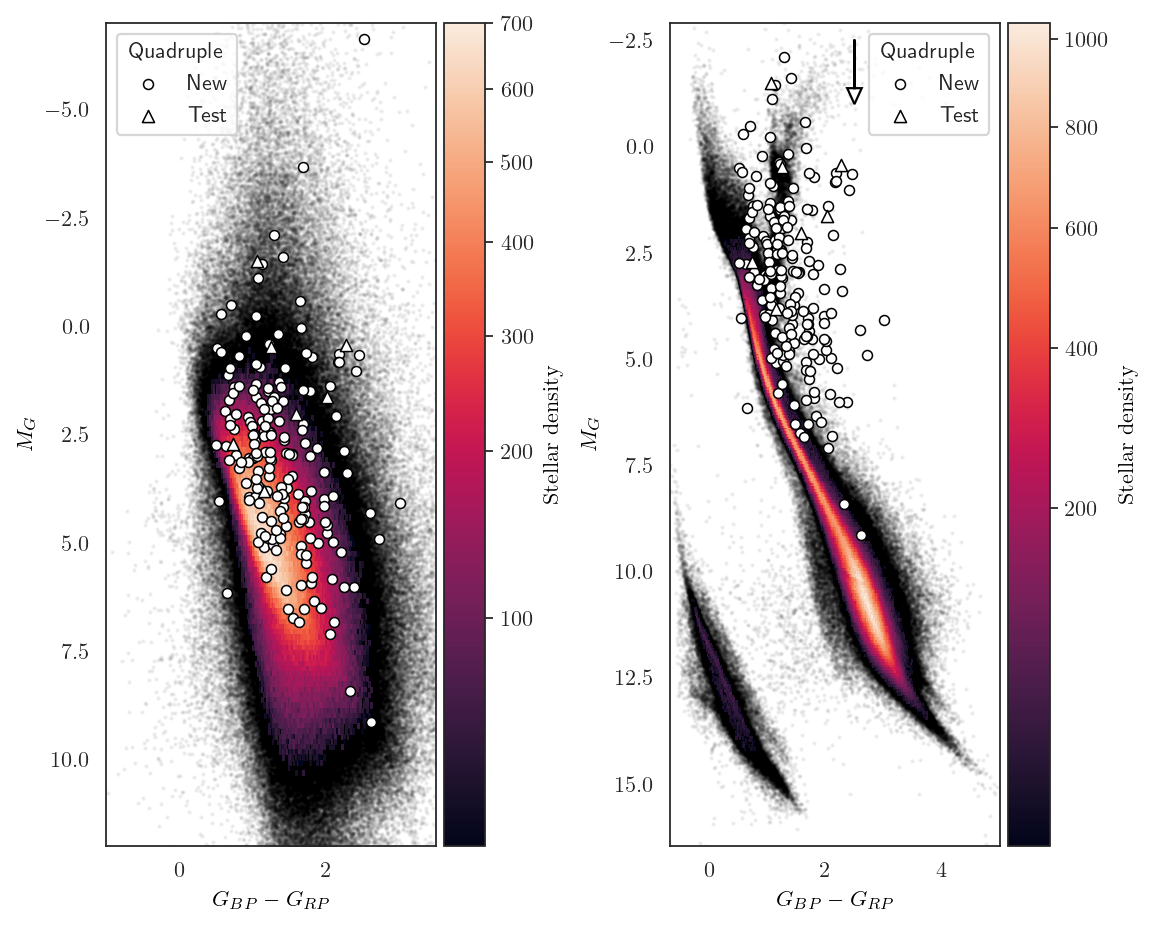}
\caption{Left: HR-diagram of all 1\,210\,001 binary objects in the sample. The double eclipsing binary candidates and test objects, separately plotted as dots, are scattered throughout the diagram and do not seem to have any preferred location. Using the sun as a reference with $ M_{G} = 5$ and $G_{BP} - G_{RP} = 1$, a larger number of objects seem to be similar to the sun.
Right: An HR-diagram of all stars in the Gaia dr3 archive within 200 pc of the Sun. The candidates and test objects are systematically above the single stars in this diagram and are mostly sun-like stars. The arrow with a length of 1.505 magnitude indicates where one star in the quadruple system would lie, assuming all stars in the system are equally bright.}
\label{fig:hr subplots}
\end{figure*}

To understand if there is any bias in our search, all 1\,210\,001 binaries in the sample were plotted in an HR-diagram, see the left panel of Fig. \ref{fig:hr subplots}. Here we see that the double eclipsing binaries, plotted separately as dots, are scattered throughout the diagram and do not seem to have any preferred location.  

Since it is interesting to not only compare the double eclipsing binaries to the input sample but also to a general star population within the Milky Way, we refer to the HR-diagram of all stars in the Gaia dr3 archive within 200 pc of the sun. This can be seen in the right panel of Fig. \ref{fig:hr subplots}. The quadruple candidates and test objects are systematically above the single stars in this diagram and are mostly Sun-like stars. We note, however, that quadruple systems on the HR-diagram have a lower magnitude of   $2.5\log_{10}(4) = 1.505$ compared to single stars, assuming all stars in the quadruple are similar and equally bright. 

\section{Discussion}\label{sec:discussion}
\subsection{Confusion with triple eclipsing systems}
As we already noted; \textit{Gaia} dr3 source ID 2003428628134264448 was previously identified as a triply eclipsing triple system \citep{Borkovits_2021} (a multistar system where all three stars eclipse each other). We initially identified this system as a quadruple because we detected a primary and secondary eclipse at a period of 88.8 days. However, as is shown by \citet{Borkovits_2021} using well-sampled TESS and ground-based follow-up \replaced{light curves}{lightcurves}; the eclipse shapes are complex and only consistent with a tertiary star that eclipses, and is being eclipsed, by a short period binary. Triple eclipsing stars are also extremely interesting for reasons similar to those of quadruple systems. For recent work on triple eclipsing systems, see \citet{Carter2011,Derekas2011,Hajdu2017,Mitnyan2020,Borkovits_2021,Powell2022,Rappaport2022,Rappaport2023, Czavalinga2023}

The fact that one of our double eclipsing binary candidates turned out to be a triply eclipsing triple system, prompted us to reconsider the long period systems that we recovered. We focus on long period systems since, for triple systems, the outer orbit has to be at least 5 times the inner orbit, while a factor of 10 is required for slightly eccentric orbits \citep{mardling_tidal_2001, Borkovits_2021}. With just the ZTF data, individual eclipses are not well-resolved, which makes an unambiguous identification of triples difficult. However, an eclipse of a triple eclipsing system is more noisy in a folded light curve because of the complex and changing shape of individual eclipses. We use this to determine whether some systems might be triple-eclipsing systems instead of double-eclipsing binaries.

Objects 2020970648976206208, 2210495392378660864, and 2038100043701450112 show somewhat irregular, long-duration eclipses in ZTF data and could be eclipsing triple systems. 
Objects 2058085351143518464, 2003428628134264448, 2201723866572302976, and 413930359370384384 show a primary and a shallow secondary eclipse at period B, with the primary eclipse somewhat irregular, suggesting that they could also be triply eclipsing systems. Object 2041659197183314304 shows well-behaved eclipses, and we judge that this system is unlikely to be a triply eclipsing system.

Objects 207943285475761280 and 2032118077650924672 are two systems for which the eclipse of the long period signal is much deeper in $g$ than $r$ and maybe a shallow secondary eclipse can be seen. This suggests that a large, cold star is eclipsed by a smaller, hotter star. We suspect that these are double eclipsing systems where one of the components has evolved off the main-sequence. For 260945106052343296 and 2012994482376100736 the period is too uncertain to draw any conclusions.

Since we cannot definitively classify these objects as triple eclipsing systems, we consider all these objects double eclipsing binary candidates in the rest of this paper. Follow-up light curve observations of the eclipses are needed for each object to definitively determine their nature.

\subsection{Orbital period distribution}
In general, binaries with shorter periods seem easier to detect than binaries with longer periods. One explanation is that stars in binaries with longer orbital periods are further apart from each other. As the orbital distance increases, the eclipse probability \begin{equation}
    \text{Eclipse probability} = \frac{R_{1}+R_{2}}{a} \propto P^{-2/3}
\end{equation}decreases since this requires a more precise alignment with the observer for the eclipse to be detected. 
In addition, longer orbital periods thus lead to shorter eclipse durations as fraction of the orbit. Some light curves showed that the eclipse duration of binary B was either very short (< 0.01 in phase), or contained very few data points, which makes it difficult for the algorithm to identify the eclipse. This agrees with the results shown in Fig. \ref{fig: jointplot periods}, where most double eclipsing binary candidates have both periods shorter than a few days.

\begin{figure}
    \centering
    \includegraphics[width=\columnwidth]{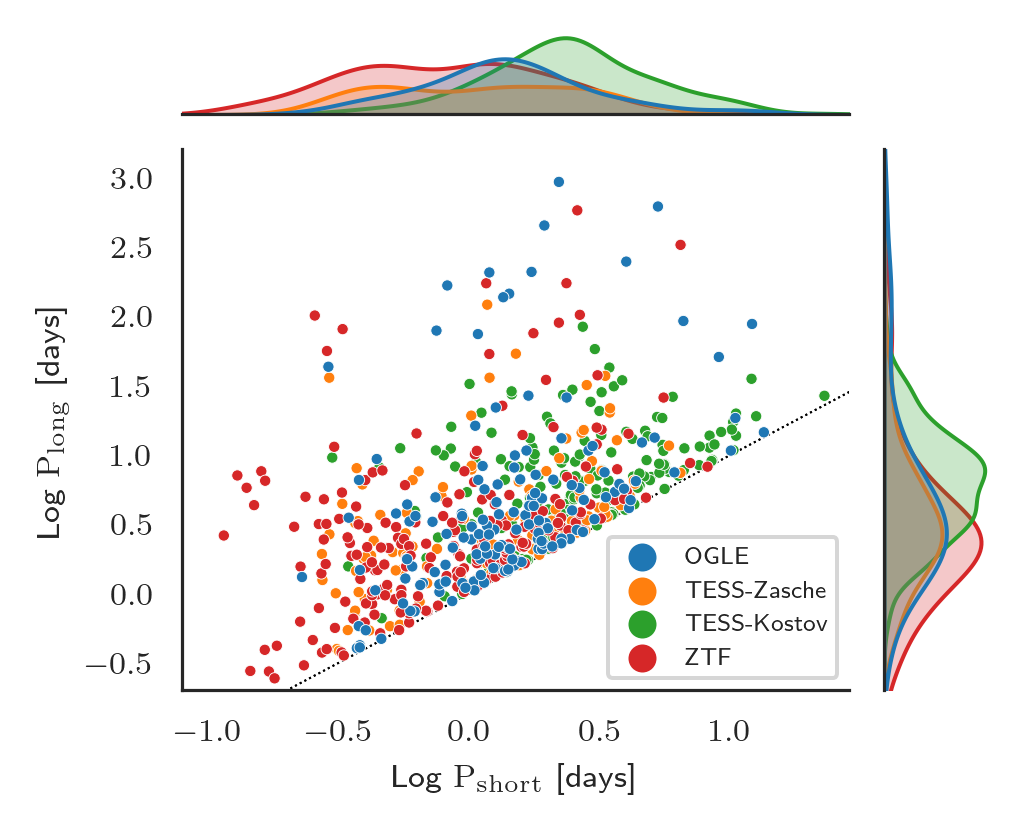}
    \caption{The orbital periods of double eclipsing binaries of the three largest samples:
    the sample from OGLE \citep{Zasche2019}, the sample from TESS by \citet{Kostov_101_2022,kostov_97_2022} and \citet{zasche_multiply_2022} and the ZTF sample from this work. Here $P_{short}$ is defined as the shortest period of A and B and $P_{long}$ as the longest period of A and B}
    \label{fig:jointplot_compared}
\end{figure}

\begin{figure}[h!] 
\centering
\includegraphics[width=\columnwidth]{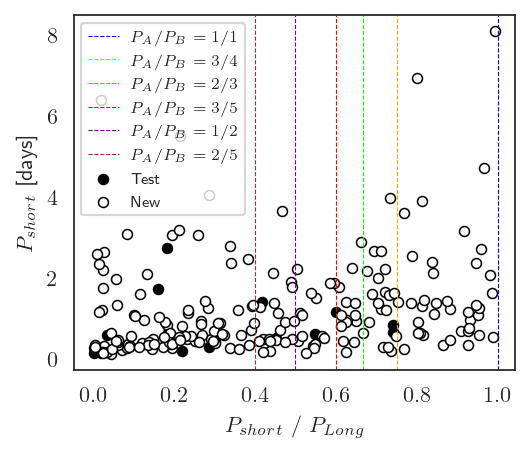}
\caption{Period ratios $\mathrm{P_{short}} / \mathrm{P_{long}}$ of each object. Here $P_{short}$ is defined as the shortest period of A and B and $P_{long}$ as the longest period of A and B. The vertical lines indicate where potential harmonics would fall. We note that integer multiple harmonics were removed in the object search. However, non integer harmonics were not removed. Here we see that most objects do not fall on any of the harmonics lines indicating that there does not seem to be any particular correlation between the periods.}
\label{fig:periods kostov harmonic}
\end{figure}

The orbital periods of the candidates found in this research seem fairly similar to those found by \citet{kostov_97_2022,zasche_multiply_2022}, see Fig. \ref{fig:jointplot_compared}. However, one difference is that we find some shorter periods in this research. Whereas previous research shows that most objects show a period ratio of $\mathrm{P_{B}}/\mathrm{P_{A}}$ between 1 and 2 \citep{kostov_97_2022}, in Fig. \ref{fig:periods kostov harmonic} we see that here also much smaller ratios are found, indicating a larger difference between period A and period B. An explanation for this can be the data cadence of TESS which is 26 days, meaning it most likely does not find any significant periods larger than approximately 13 days. Furthermore, short 30 minute exposures by TESS may lead to the spreading out of short eclipses, which is why TESS might not find very short periods either. We also note that all stars in our sample are listed in the \textit{Gaia} catalogue. Any bias of \textit{Gaia} towards shorter periods may therefore reflect in our results. 

\subsection{Detection efficiency and completeness}
In this section, we briefly discuss the detection efficiency and completeness of our search. Of the 575\,526  analysed objects, at least 203 ($\approx$ 0.035\%) passed the visual inspection tests and showed to be a double eclipsing binary candidate rather than a binary system. This is approximately a factor 4.5 lower success rate than \citet{zasche_multiply_2022} (116 out of 70.000) and a factor 7 lower than \citet{pawlak_eclipsing_2013} (15 out of 6138). 
However, we note that OGLE-III has observed all stars in a dense area in the Small Magellanic Cloud for a longer period of time \citep{pawlak_eclipsing_2013}. \citet{zasche_multiply_2022} have, like this research, only looked at eclipsing binaries, but a difference with this research is the data cadence, which for ZTF typically is one observation per night only for five years. TESS, on the other hand, provides continuous undisturbed photometry for 26 days \citep{zasche_multiply_2022}. Lastly, the angular resolution of ZTF, OGLE, and TESS are very different.
This, combined with the new method used here, makes it difficult to compare success rates, but could explain the difference. 

In the rest of this section, we briefly discuss a few causes that limit the completeness of our search.
First, we consider the efficiency of our quadruple detection method. Our method was able to retrieve the correct periods for most known objects in the test dataset. For periods that could not be recovered, further investigation revealed why this was not possible. In some cases, many data points were flagged as unreliable, leaving only very few data points in the light curve for period finding. In other instances, the eclipse depth of one binary was very small, on the order of a few percent, making it difficult for the algorithm to detect any periodic signal. Other reasons why the algorithm was not successful included that some light curves showed a very small trend in flux over time, resulting in the cadence of the telescope being found as significant period; this typically being one or two days. Even light from the Moon coming in as background noise could lead to the period of the Moon being found to be a significant period. For compact double eclipsing binaries, eclipse-timing variations (ETV) can smear periodic signals which may affect the period finding. However, when comparing our results with known systems with ETV, we found this not to be the case. We therefore expect this method to be suitable for finding compact double eclipsing binaries.

The completeness of our search for double eclipsing binaries is also limited by our use of the \textit{Gaia} eclipsing binary catalogue.
When we look at the results of other studies \citep{kostov_97_2022, Kostov_101_2022, zasche_multiply_2022}, we see that approximately 30\% of the found quadruples are in the \textit{Gaia} catalogue. In this research we have only looked at ZTF objects that are known in the \textit{Gaia} catalogue. This means that potentially three times more quadruples can be found in the ZTF data that are not in the \textit{Gaia} catalogue.  

As it was not practical to look at 575\,526 light curves, we used automatically calculated statistics to significantly decrease the number of light curves to inspect which could also affect the completeness of our search of the ZTF data. 
We used the test objects to empirically determine cuts; relative peak height $>10$ and S/N $>10$. As can be seen in Fig. \ref{fig:RPH and SNR all objects}, the wide range of values of relative peak height and S/N cannot reliably indicate whether an object might be a quadruple candidate. This suggests that more double eclipsing binaries can be found for values of relative peak height and S/N below 10. It can also be seen in the figure that some objects did have a high relative peak height and S/N but were not identified as a quadruple candidate. This could have several reasons.
In some cases, the algorithm was not able to completely remove the first periodic signal, which means that the signal, or an harmonic, was returned again for period B. Other explanations include, but are not limited to, an outlier in the light curve, the cadence of the telescope, or the influence of the Moon. 
Sometimes the algorithm did not find an eclipsing signal but a sinusoidal signal, possibly due to the pulsation of one of the stars in the eclipsing binary. Even though this is not a double eclipsing binary, a strong periodic signal was still found, resulting in a high relative peak height and S/N. 

\begin{figure*}[h!] 
\centering
\includegraphics[width=\textwidth]{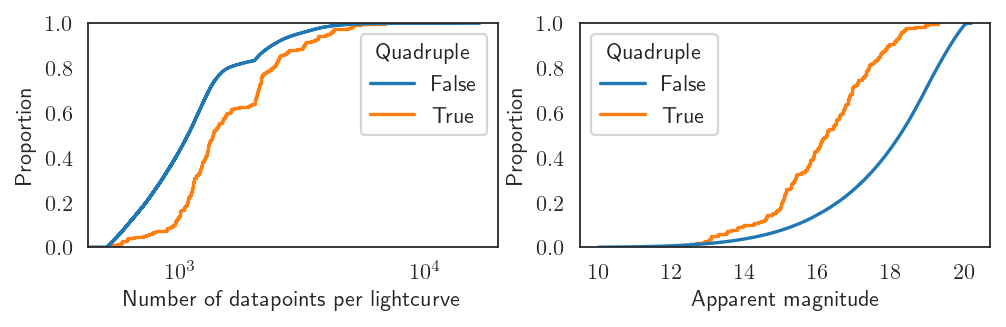}
\caption{Left: Cumulative histogram for the number of data points per light curve. After passing a threshold of approximately 1000 data points there is a significant increase in the quadruple detection efficiency. Note that the plateau and discontinuity at around 2000 data points is a result of the ZTF survey strategy. Right: Cumulative histogram for the magnitude of stars for the complete sample and quadruple candidates in the magnitude range 13-20.5. We see that for stars fainter than $G\gtrsim18$ the quadruple detection efficiency rapidly approaches zero. The sharp increase at $G\approx15$ is not fully understood.}
\label{fig: histogram}
\end{figure*}

We suspected that the detection efficiency of double eclipsing binary stars is also strongly affected by the number of epochs in the \replaced{light curve}{ligthcurve}.
If we look at the left panel in Fig. \ref{fig: histogram} we see a cumulative histogram for the number of data points per light curve. It shows that, only after passing a threshold of approximately 1000 data points, the quadruple detection efficiency significantly increases. After that, there are more double eclipsing binaries as the number of data points increases. The difference, however, is small.
This means that, as ZTF keeps collecting data (or ZTF is combined with external data), we can expect the number of detectable double eclipsing binaries to double.

We also briefly consider to what brightness we can reliably find double eclipsing binaries in ZTF data.
On the right panel in Fig. \ref{fig: histogram} we see a cumulative histogram for the magnitude of stars, both for the complete sample and for the quadruple candidates. The ZTF has looked at stars in the magnitude range 13-20.5. 
Since the curve of the quadruples is constantly above the overall sample curve, this indicates that it is easier to find quadruples for bright stars compared to faint stars. Almost all quadruples have a magnitude of $G\lesssim18$, while only half of the input sample is fainter than this limit. We conclude that the light curve signal-to-noise ratio is too low for sources fainter than $G\approx18$. As can be seen in \citet{masci2019}, the RMS precision is $\approx 1\%$ for sources brighter than 18 magnitude, but the precision degrades rapidly for fainter sources.
 
To summarise, we expect that future research can find at least three to four times more quadruples by including the following considerations. First, the search should be not limited to the \textit{Gaia} eclipsing binary catalogue but the entire ZTF database can be searched instead. Secondly, systems with a relative peak height and S/N lower than 10 should also be considered. Thirdly, longer observation time of the ZTF will lead to more data points which has a positive effect on quadruple finding. 
Lastly, by using an automatic detection method that can replace visual detection, for instance machine learning, all the light curves can be analysed more efficiently.

\section{Conclusions and future work}\label{sec:conlusion}

We have developed a method to systematically search for double eclipsing binaries in sparsely sampled data. Using this method we found 198 new double eclipsing binaries candidates in the ZTF data.
The periods of the objects ranged from 0.11 days to 323 days with a mean of periods A and B of 1.21 days and 10.52 days respectively. 
By searching the entire ZTF database, increasing the number of data points per light curve, and implementing an automatic detection mechanism that replaces visual inspection, we expect that at least three to four times more quadruples can be found using this method. Other recommendations for future work are; to apply this method to data collected by other telescopes such as \textit{Gaia}. In addition, we recommend that the multi star systems are included in the target lists of large multiplex spectroscopic surveys such as SDSS, DESI, and WEAVE in order to obtain phase-resolved spectra of the quadruple candidates to further investigate their properties. Lastly, detection of eclipse-timing variations in the double eclipsing binaries can definitively prove their quadruple nature.

\begin{acknowledgements}
We thank Silvia Toonen for a useful discussion about multi-star systems.

This publication is part of the project "The life and death of white dwarf binary stars" (with project number VI.Veni.212.201) of the research programme NWO Talent Programme Veni Science domain 2021 which is financed by the Dutch Research Council (NWO).

Based on observations obtained with the Samuel Oschin 48-inch Telescope at the Palomar Observatory as part of the Zwicky Transient Facility project. ZTF is supported by the National Science Foundation under Grant No. AST-1440341 and a collaboration including Caltech, IPAC, the Weizmann Institute for Science, the Oskar Klein Center at Stockholm University, the University of Maryland, the University of Washington, Deutsches Elektronen-Synchrotron and Humboldt University, Los Alamos National Laboratories, the TANGO Consortium of Taiwan, the University of Wisconsin at Milwaukee, and Lawrence Berkeley National Laboratories. Operations are conducted by COO, IPAC, and UW.

Based on observations obtained with the Samuel Oschin Telescope 48-inch and the 60-inch Telescope at the Palomar Observatory as part of the Zwicky Transient Facility project. ZTF is supported by the National Science Foundation under Grants No. AST-1440341 and AST-2034437 and a collaboration including current partners Caltech, IPAC, the Weizmann Institute for Science, the Oskar Klein Center at Stockholm University, the University of Maryland, Deutsches Elektronen-Synchrotron and Humboldt University, the TANGO Consortium of Taiwan, the University of Wisconsin at Milwaukee, Trinity College Dublin, Lawrence Livermore National Laboratories, IN2P3, University of Warwick, Ruhr University Bochum, Northwestern University and former partners the University of Washington, Los Alamos National Laboratories, and Lawrence Berkeley National Laboratories. Operations are conducted by COO, IPAC, and UW.

The ztfquery code was funded by the European Research Council (ERC) under the European Union's Horizon 2020 research and innovation programme (grant agreement n°759194 - USNAC, PI: Rigault).

This research has made use of the VizieR catalogue access tool, CDS, Strasbourg, France. This research made use of NumPy \citep{harris2020array} This research made use of matplotlib, a Python library for publication quality graphics \citep{Hunter:2007} This research made use of Astroquery \citep{2019AJ....157...98G} This research made use of Astropy, a community-developed core Python package for Astronomy \citep{astropy:2022} 

\end{acknowledgements}

%-------------------------------------------------------------------

\bibliographystyle{aa} % style aa.bst
\bibliography{biblio}
%\begin{thebibliography}
%\end{thebibliography}

\begin{appendix} %Second online appendix
\section{Additional figures}
\begin{figure}[h!] 
\centering
\includegraphics[width=\columnwidth]{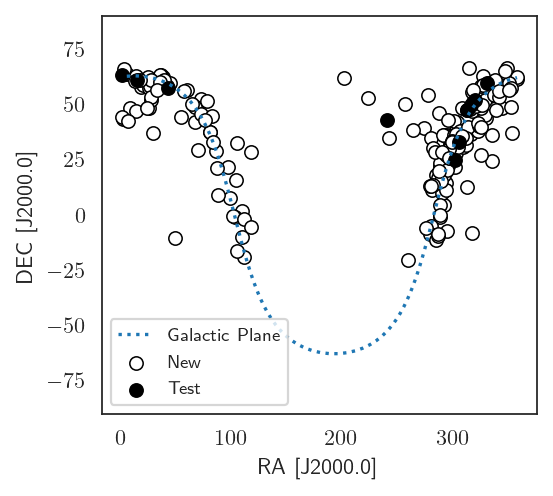}
\caption{Right ascension (RA) and declination (DEC) of the double eclipsing binary candidates and test objects in this research. As expected, most candidates lie in the Galactic plane. No objects are seen below a declination of -30\degree\ as this is not visible by the ZTF.}
\label{fig:location diagram}
\end{figure}

For the ZTF, the periods of objects that lie in the Galactic plane can be more easily retrieved. This is can also be seen in Fig. \ref{fig:location diagram} where the location of all quadruple candidates and test objects are plotted. These results agree with the results found in earlier research \citep{zasche_multiply_2022, kostov_97_2022}. It is most likely that the number of double eclipsing binaries is higher in the Galactic plane as a result of higher star density.

\begin{figure*}[] 
\includegraphics[width=\textwidth]{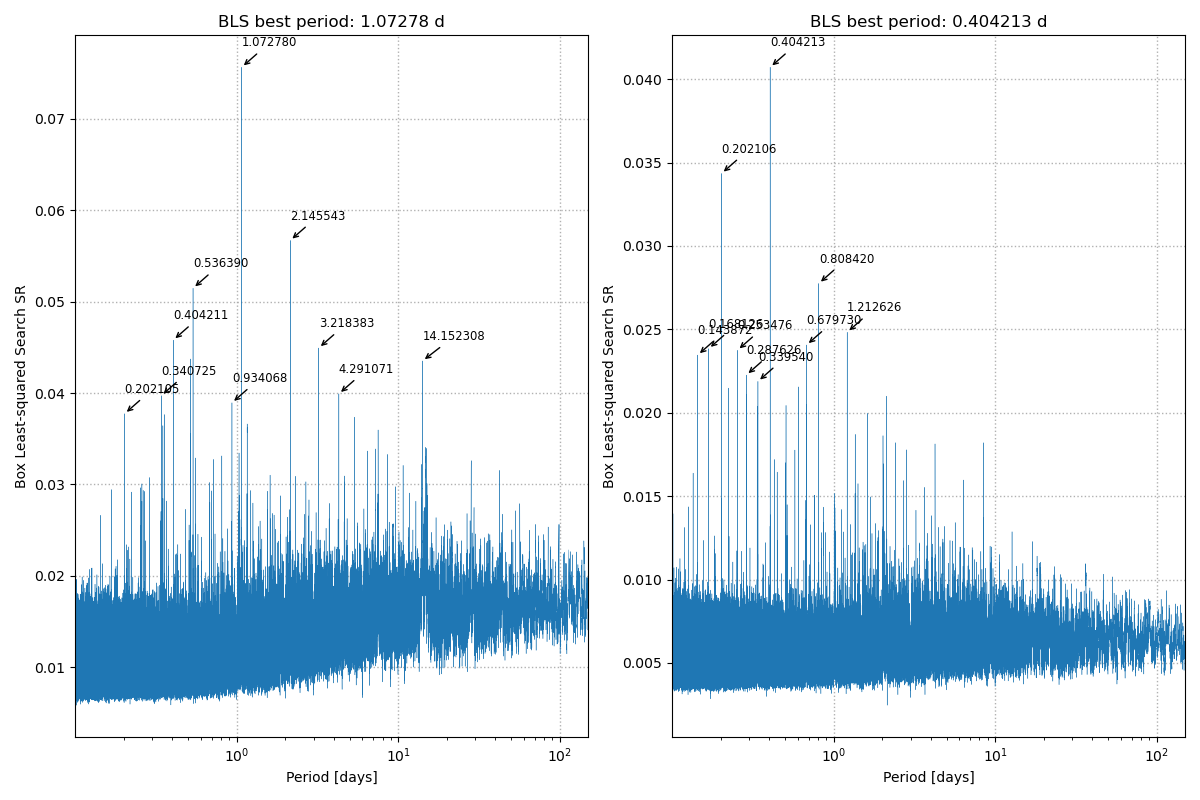}
\caption{A periodogram of Gaia dr3 source ID 2216420454386370176 showing the lspvals as a function of orbital period. Left: The periodogram power values for period A. Right: The periodgram power values for period B after removing the signal of period A.}
\label{fig:periodogram}
\end{figure*}

\begin{figure*}[] 
\includegraphics[width=\textwidth]{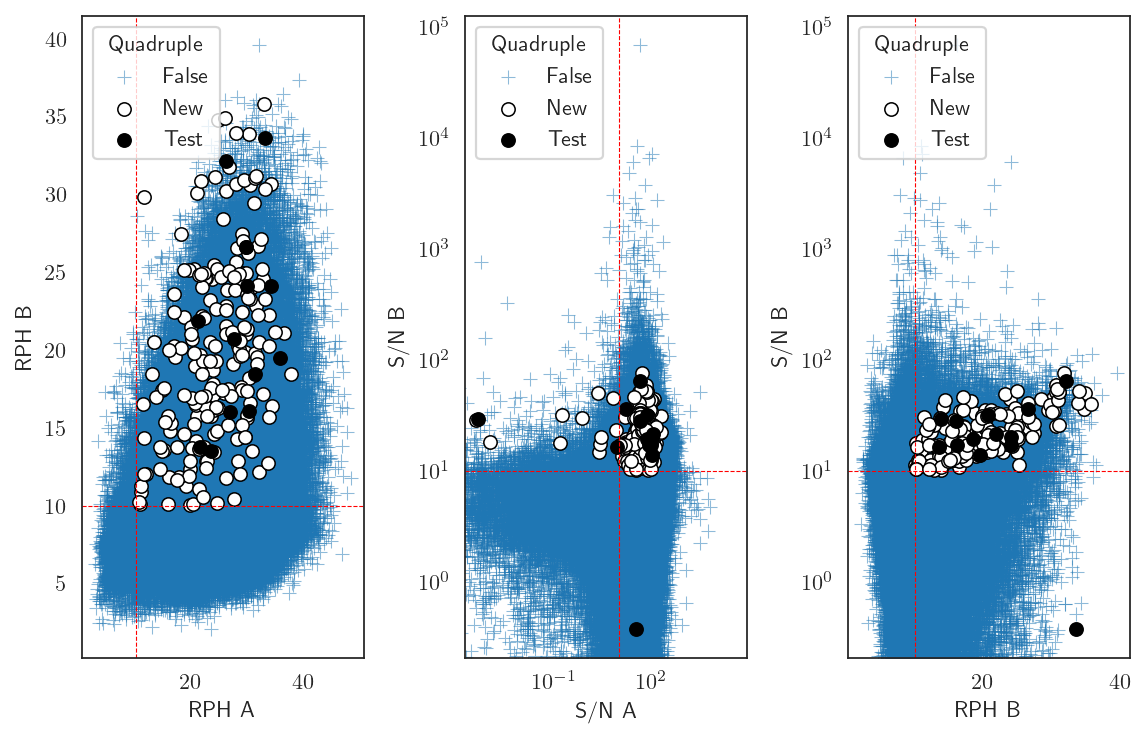}
\caption{The relative peak height and S/N of all objects including double eclipsing binary candidates and test objects.
Left: Relative peak heights A and B for all objects. Middle: S/N A and S/N B results for all objects. Right: S/N B and relative peak height B of all objects. We see that quadruples can have a wide range of RPH and S/N values. However, this also means that more can be found at values of RPH < 10 \& S/N <10. }
\label{fig:RPH and SNR all objects}
\end{figure*}

\clearpage
\section{Summary of double eclipsing binary candidates}
Table \ref{tab:longtable} in this appendix lists all new double eclipsing binaries we found in ZTF.
\end{appendix}

% \listofobjects
\begin{table*}
\centering

    \small
    \caption{Summary of the double eclipsing binary candidates. For each candidate the location (RA \& DEC), Gaia dr3 source ID, periods ($\mathrm{P_{A}[d]}$ \& $\mathrm{P_{B}[d]}$) and eclipse depths ($\mathrm{D_{A}}$ \& $\mathrm{D_{B}}$) are given. Four objects have been found in previous research and are marked here with an asterisk. The object marked with a "T" was initially identified as a quadruple, but further investigation revealed that this is, in fact, a triple system. } 
    \label{tab:longtable}
\begin{tabular}{ccccccc}
RA {[}J2000.0{]} & DEC {[}J2000.0{]} & Gaia dr3 source ID & $\mathrm{P_{A} [d]}$ & $\mathrm{D_{A}}$ & $\mathrm{P_{B} [d]}$ & $\mathrm{D_{B}}$ \\ 
\hline \hline
1.461502 & 44.447714 & 386474080852740992 & 2.599973 & 0.178 & 0.114405 & 0.017  \\
2.702797 & 43.426465 & 384686137507666944 & 0.561156 & 0.057 & 0.30431 & 0.02  \\
3.255083 & 66.065778 & 528173641885393792 & 1.377078 & 0.007 & 1.958498 & 0.067  \\
5.530266 & 63.915121 & 431071367690614016 & 1.993434 & 0.207 & 2.83458 & 0.12  \\
6.850741 & 42.275447 & 382118159381427840 & 0.302332 & 0.045 & 11.331499 & 0.063  \\
8.806599 & 48.41824 & 390942874067575936 & 0.254808 & 0.066 & 100.03476 & 0.159  \\
13.007138 & 60.631402 & 427296611137250432 & 1.39752 & 0.108 & 1.233086 & 0.034  \\
13.938753 & 62.946494 & 523702890174595456 & 0.620434 & 0.129 & 0.766964 & 0.071  \\
14.457916 & 46.742013 & 377967807928167552 & 1.557556 & 0.059 & 0.569134 & 0.181  \\
17.516555 & 60.901287 & 522445255030705920 & 0.357662 & 0.183 & 0.653228 & 0.098  \\
18.801672 & 57.606248 & 413446544200880896 & 0.951178 & 0.076 & 7.547427 & 0.073  \\
20.433859 & 58.575703 & 413930359370384384 & 1.745764 & 0.211 & 74.595949 & 0.184  \\
23.839866 & 48.216903 & 399464707656254976 & 6.934429 & 0.026 & 8.655612 & 0.066  \\
24.594285 & 62.368258 & 511572184543165184 & 1.111992 & 0.017 & 0.893415 & 0.099  \\
25.46978 & 48.453185 & 405343761970536320 & 3.214925 & 0.181 & 0.475366 & 0.051  \\
26.105725 & 57.363963 & 505854376178706048 & 0.395542 & 0.283 & 0.613647 & 0.075  \\
26.99833 & 58.484187 & 506382008620308352 & 0.469946 & 0.117 & 1.607418 & 0.082  \\
27.345582 & 51.957608 & 407610302112852992 & 0.27613 & 0.067 & 4.742124 & 0.105  \\
27.478311 & 61.054208 & 511202954794627072 & 2.522958 & 0.088 & 2.118876 & 0.037  \\
27.619643 & 53.266914 & 407849617690288128$^*$ & 1.388514 & 0.182 & 1.842243 & 0.133  \\
29.369827 & 36.938288 & 330442487264141952 & 2.165996 & 0.148 & 3.201325 & 0.069  \\
32.660282 & 56.361212 & 457045959812411520 & 2.907525 & 0.249 & 1.766785 & 0.083  \\
35.274278 & 59.921344 & 507367789513371392 & 4.733592 & 0.119 & 1.111969 & 0.06  \\
35.979435 & 63.411351 & 514335841379134336 & 1.718536 & 0.264 & 1.220503 & 0.075  \\
36.881203 & 63.214977 & 514291963992396928 & 3.006382 & 0.093 & 0.212537 & 0.019  \\
38.031578 & 62.343023 & 513987498047116544 & 1.019208 & 0.086 & 0.948252 & 0.071  \\
38.292867 & 59.484742 & 465028723463664384 & 3.460204 & 0.192 & 0.92533 & 0.045  \\
39.195205 & 60.984813 & 465274395594682240 & 3.887904 & 0.148 & 1.896361 & 0.083  \\
42.417307 & 58.265701 & 461012963401936256 & 1.060832 & 0.048 & 10.593118 & 0.034  \\
44.301404 & 59.454303 & 461553472144934656 & 0.270063 & 0.053 & 1.385304 & 0.049  \\
49.178341 & -10.308834 & 5166225811703486464 & 0.275808 & 0.248 & 2.432133 & 0.098  \\
54.711333 & 44.047482 & 244408588611828096 & 0.178865 & 0.145 & 0.243028 & 0.022  \\
57.458651 & 56.117922 & 445534966419218560 & 1.292364 & 0.246 & 1.088451 & 0.051  \\
59.624908 & 56.534336 & 469545650369412864 & 0.224785 & 0.178 & 1.551227 & 0.05  \\
64.168902 & 50.317876 & 271381051952218496 & 2.245528 & 0.108 & 1.40276 & 0.05  \\
67.213782 & 42.115243 & 228205115816360832 & 1.125492 & 0.13 & 2.201828 & 0.02  \\
67.217417 & 48.572915 & 258090121033397248 & 5.48996 & 0.156 & 25.643685 & 0.035  \\
68.629624 & 48.698441 & 257683439171736576 & 0.81849 & 0.107 & 4.499209 & 0.08  \\
69.925315 & 29.134575 & 158123726424621824$^*$ & 1.58676 & 0.128 & 13.783777 & 0.06  \\
72.308262 & 52.265466 & 260945106052343296 & 3.070683 & 0.241 & 37.120622 & 0.061  \\
73.106774 & 45.534572 & 206454920393197440 & 0.583875 & 0.365 & 0.715742 & 0.089  \\
76.053644 & 42.807356 & 202201601392851072 & 0.624014 & 0.05 & 14.132366 & 0.087  \\
77.972018 & 51.672042 & 262779881722935424 & 1.637614 & 0.193 & 2.266959 & 0.101  \\
80.461508 & 37.37017 & 184468231186516352 & 2.345842 & 0.133 & 6.86879 & 0.09  \\
80.679986 & 37.664808 & 184497570107735296 & 2.3771 & 0.094 & 2.839768 & 0.055  \\
82.738969 & 44.803406 & 207943285475761280 & 52.829036 & 0.274 & 1.184305 & 0.066  \\
83.858615 & 33.004921 & 3448886540015107584 & 2.989964 & 0.115 & 0.52143 & 0.06  \\
86.591414 & 28.923839 & 3443402279093860096 & 0.53534 & 0.144 & 0.541528 & 0.079  \\
87.5282 & 21.330875 & 3400120484904001408 & 3.572446 & 0.166 & 1.759766 & 0.066  \\
88.00886 & 8.805691 & 3335190578070549632 & 0.36877 & 0.083 & 1.890454 & 0.068  \\
96.842245 & 21.72127 & 3376140681762631168 & 0.39992 & 0.093 & 0.842531 & 0.049  \\
98.587002 & 7.745466 & 3325963098535865728 & 0.584642 & 0.307 & 0.6136 & 0.053  \\
\hline
\end{tabular}
\end{table*}

\begin{table*}
\centering

\ContinuedFloat
    \small
    \caption{Continued}

\begin{tabular}{ccccccc}
RA {[}J2000.0{]} & DEC {[}J2000.0{]} & Gaia dr3 source ID & $\mathrm{P_{A} [d]}$ & $\mathrm{D_{A}}$ & $\mathrm{P_{B} [d]}$ & $\mathrm{D_{B}}$ \\ 
\hline \hline
101.872246 & -0.444575 & 3113400013097881600 & 0.86539 & 0.121 & 2.182992 & 0.048  \\
103.057739 & -1.125556 & 3112366227355971328 & 0.948312 & 0.155 & 0.632336 & 0.069  \\
104.069404 & 15.809035 & 3355042672829484032 & 2.502663 & 0.177 & 0.530042 & 0.06  \\
105.184549 & 32.479562 & 890716293806690816 & 0.263998 & 0.105 & 3.144128 & 0.136  \\
105.242128 & -16.522209 & 2935716391431248256 & 0.424388 & 0.076 & 7.412474 & 0.067  \\
109.571295 & 1.765508 & 3111928351141425792 & 0.274758 & 0.174 & 0.144463 & 0.054  \\
109.612639 & -10.149447 & 3047190824491991168 & 0.45911 & 0.163 & 0.25037 & 0.05  \\
111.089388 & -2.032067 & 3061770932787038336 & 0.422212 & 0.082 & 0.966995 & 0.054  \\
111.47435 & -19.281677 & 2930398740886024064 & 0.347152 & 0.219 & 2.294894 & 0.144  \\
117.697495 & 28.467895 & 875709132614498560 & 0.442358 & 0.158 & 1.466414 & 0.069  \\
118.096034 & -5.697484 & 3044312990239139968 & 1.452632 & 0.176 & 1.779009 & 0.072  \\
202.102553 & 61.876817 & 1663741485747221248 & 2.770754 & 0.15 & 8.163296 & 0.08  \\
223.42518 & 52.715832 & 1594082407606370176 & 2.700472 & 0.074 & 1.488501 & 0.043  \\
242.649013 & 34.62108 & 1323756753679794048 & 7.041158 & 0.114 & 0.128904 & 0.014  \\
257.277242 & 49.948796 & 1414515322519168640 & 0.301544 & 0.197 & 0.231617 & 0.019  \\
259.250732 & -20.297582 & 4115999620890627968 & 0.621141 & 0.24 & 0.773402 & 0.114  \\
263.775976 & 38.311866 & 1343353448205456256 & 3.048348 & 0.095 & 11.792556 & 0.033  \\
273.743299 & 39.207529 & 2109307852667550464 & 2.205617 & 0.187 & 4.373204 & 0.029  \\
276.008132 & -5.774128 & 4161007815803011968 & 3.070536 & 0.211 & 1.371504 & 0.102  \\
278.026086 & 54.402996 & 2147458221793949184 & 3.109478 & 0.097 & 1.853905 & 0.079  \\
279.117985 & -8.525243 & 4156431820198005376 & 1.407928 & 0.281 & 1.331908 & 0.085  \\
279.719282 & 13.163625 & 4508288602094506112 & 0.399492 & 0.061 & 1.528379 & 0.102  \\
280.635591 & -5.252461 & 4256607950976483968 & 1.907376 & 0.222 & 0.903688 & 0.036  \\
280.635808 & 13.120684 & 4505600296166989312 & 4.870614 & 0.097 & 4.70215 & 0.019  \\
280.713417 & 34.65467 & 2091761983553949696 & 0.271486 & 0.204 & 0.372731 & 0.028  \\
281.121588 & 11.13593 & 4504043245957168640 & 10.015578 & 0.077 & 1.042907 & 0.022  \\
282.145741 & 30.399305 & 2041659197183314304 & 0.283664 & 0.061 & 55.568657 & 0.06  \\
282.312661 & 13.469479 & 4505497904155844992 & 0.28221 & 0.303 & 3.150768 & 0.103  \\
283.996449 & 28.282368 & 2040256804464235648 & 5.739774 & 0.104 & 0.139815 & 0.012  \\
284.492451 & -11.193995 & 4201927347951103488 & 0.513114 & 0.405 & 0.453093 & 0.05  \\
285.268359 & 18.064422 & 4517289341731165568 & 1.056042 & 0.07 & 0.545708 & 0.031  \\
286.391765 & 16.558561 & 4513882917294832384 & 0.875846 & 0.043 & 2.838248 & 0.022  \\
286.904493 & -9.457319 & 4204113619406564480 & 0.33327 & 0.097 & 0.867417 & 0.047  \\
287.114664 & -8.543055 & 4204326997705570304 & 0.272314 & 0.3 & 0.170204 & 0.061  \\
287.445354 & 19.597742 & 4516377056329377664 & 1.035768 & 0.171 & 6.325886 & 0.082  \\
287.648865 & 16.355321 & 4513263857861420672 & 0.45636 & 0.166 & 2.129335 & 0.1  \\
287.884621 & 46.028573 & 2130268255146656128 & 0.422266 & 0.269 & 1.029215 & 0.035  \\
288.026347 & 23.298684 & 4521240642906912128 & 3.00593 & 0.011 & 1.616295 & 0.071  \\
288.248394 & 0.080909 & 4264206469687918848 & 3.166976 & 0.208 & 1.852653 & 0.103  \\
288.471817 & -1.094539 & 4263034768239285632 & 1.862082 & 0.16 & 0.592354 & 0.053  \\
288.648651 & 22.157332 & 4520164564651728128 & 0.355088 & 0.09 & 0.32902 & 0.045  \\
288.825262 & 4.312846 & 4293109022582320640 & 0.390062 & 0.277 & 0.164066 & 0.11  \\
289.517888 & 15.160579 & 4320727174167431296 & 2.163692 & 0.033 & 1.607176 & 0.042  \\
290.184501 & 28.541396 & 2038100043701450112 & 0.325552 & 0.122 & 79.87107 & 0.122  \\
290.604938 & 9.66816 & 4308597705466532224 & 5.097978 & 0.173 & 1.412602 & 0.057  \\
290.991691 & 35.958626 & 2050141310916611584 & 3.877374 & 0.021 & 4.766349 & 0.047  \\
291.640789 & 4.331223 & 4292639977810529152 & 1.64215 & 0.063 & 1.061804 & 0.058  \\
291.714347 & 18.228635 & 4323257219102940288 & 1.588804 & 0.186 & 2.302579 & 0.077  \\
291.820676 & 35.97582 & 2049954600104034432 & 4.030792 & 0.113 & 14.062535 & 0.062  \\
293.462031 & 19.846805 & 1825483906846429056 & 6.45557 & 0.145 & 2.467705 & 0.047  \\
293.504188 & 27.432478 & 2025279046615280512 & 2.759184 & 0.134 & 0.863998 & 0.026  \\
293.592507 & 11.006839 & 4314657904341880960 & 3.349092 & 0.109 & 1.32064 & 0.027  \\
293.657387 & 19.881442 & 1825578567909317760 & 4.770348 & 0.152 & 2.118271 & 0.038  \\
293.823697 & 25.752582 & 2021549301357843328 & 0.417266 & 0.168 & 0.182446 & 0.089  \\
\hline
\end{tabular}
\end{table*}

\begin{table*}
\centering

\ContinuedFloat
    \small
    \caption{Continued}

\begin{tabular}{ccccccc}
RA {[}J2000.0{]} & DEC {[}J2000.0{]} & Gaia dr3 source ID & $\mathrm{P_{A} [d]}$ & $\mathrm{D_{A}}$ & $\mathrm{P_{B} [d]}$ & $\mathrm{D_{B}}$ \\ 
\hline \hline
294.156333 & 14.316215 & 4318042922678224768 & 1.840298 & 0.023 & 0.915486 & 0.087  \\
294.332664 & 27.158424 & 2025146250574308480 & 0.430934 & 0.184 & 0.698726 & 0.062  \\
294.518815 & -7.404433 & 4207090654517833344 & 3.426032 & 0.186 & 0.315937 & 0.04  \\
294.832597 & 26.871716 & 2025081478136260480 & 3.072568 & 0.066 & 2.212912 & 0.052  \\
294.839444 & 35.975154 & 2048268017993738624 & 0.234828 & 0.043 & 4.944031 & 0.052  \\
294.933937 & 20.202009 & 1825745006505362048 & 7.807304 & 0.051 & 3.649778 & 0.042  \\
295.309706 & 42.822037 & 2077913565886105344 & 0.373692 & 0.344 & 0.322663 & 0.03  \\
295.479516 & 34.635703 & 2047323949824485376 & 2.346872 & 0.18 & 2.481432 & 0.055  \\
295.527859 & 22.054362 & 1827708802964917888 & 0.545902 & 0.187 & 0.96728 & 0.047  \\
295.87929 & 25.881643 & 2021774220180193536 & 1.357318 & 0.0 & 2.09345 & 0.026  \\
296.235173 & 29.583425 & 2031977619399032448 & 8.147016 & 0.064 & 8.083514 & 0.023  \\
296.585034 & 23.395937 & 2020056405032433152 & 0.812118 & 0.18 & 0.753871 & 0.086  \\
296.895517 & 25.636448 & 2020970648976206208 & 6.375142 & 0.168 & 323.201568 & 0.109  \\
296.955946 & 30.241753 & 2032118077650924672 & 171.076078 & 0.28 & 2.335024 & 0.089  \\
297.00612 & 24.058537 & 2020490437267187712 & 1.35866 & 0.148 & 1.729201 & 0.093  \\
297.034812 & 30.08806 & 2031925182082532608 & 3.590404 & 0.001 & 0.789163 & 0.032  \\
297.056926 & 25.787105 & 2026976864358731904 & 1.425158 & 0.0 & 0.89166 & 0.105  \\
298.002327 & 31.866931 & 2033828574151981184 & 1.768354 & 0.231 & 0.571495 & 0.122  \\
298.172845 & 26.864218 & 2027154955201119488 & 1.647592 & 0.093 & 1.191313 & 0.026  \\
298.530047 & 24.682368 & 1834469768675202560 & 2.881026 & 0.081 & 4.353022 & 0.036  \\
298.820248 & 32.01454 & 2033656156987971840 & 7.548416 & 0.327 & 0.159107 & 0.049  \\
298.881534 & 29.89156 & 2030312580776346240 & 0.462184 & 0.179 & 7.651529 & 0.06  \\
298.908525 & 32.746538 & 2034277144881784704 & 2.452324 & 0.245 & 0.558699 & 0.026  \\
299.522067 & 30.246542 & 2030426861313716224 & 1.305256 & 0.061 & 1.607371 & 0.019  \\
299.620393 & 33.527522 & 2034419218049672704 & 1.411226 & 0.099 & 1.614413 & 0.072  \\
299.976835 & 29.83971 & 2030212490900354176 & 1.189566 & 0.022 & 3.314311 & 0.025  \\
301.355473 & 34.989363 & 2058617549147113728 & 2.08353 & 0.045 & 15.704131 & 0.051  \\
301.37899 & 35.97206 & 2059117002318116096 & 1.741608 & 0.114 & 1.064947 & 0.033  \\
301.391172 & 42.318524 & 2074882590295669376 & 1.127028 & 0.197 & 2.46579 & 0.132  \\
302.250175 & 21.436492 & 1829670778385777920 & 5.414028 & 0.109 & 3.970943 & 0.085  \\
302.637062 & 38.400325 & 2061735248734886400 & 1.079346 & 0.226 & 0.468297 & 0.029  \\
302.98334 & 36.463072 & 2059178162649303808 & 5.993378 & 0.266 & 0.563785 & 0.056  \\
303.229266 & 27.908953 & 1836995514370597120 & 0.374674 & 0.076 & 3.126105 & 0.05  \\
303.523133 & 38.382872 & 2060949991271681664$^*$ & 1.144666 & 0.129 & 0.481849 & 0.017  \\
304.151811 & 41.930677 & 2068658632915468672 & 4.363316 & 0.078 & 1.242846 & 0.052  \\
304.756649 & 37.920034 & 2061050802729418496 & 1.325726 & 0.172 & 0.813421 & 0.097  \\
304.795015 & 30.560399 & 1861612037845705728 & 5.302642 & 0.194 & 0.323736 & 0.037  \\
306.408352 & 37.894974 & 2058085351143518464 & 1.152476 & 0.193 & 85.498447 & 0.139  \\
308.681181 & 48.20847 & 2167733048719069312 & 2.17693 & 0.17 & 0.584749 & 0.086  \\
308.779214 & 35.826622 & 2056704914305732096 & 5.152764 & 0.146 & 0.910183 & 0.069  \\
309.435623 & 30.852742 & 1862347439319475072 & 0.920364 & 0.145 & 1.417808 & 0.094  \\
311.55134 & 31.682545 & 1859824197571796736 & 0.283018 & 0.133 & 0.394832 & 0.064  \\
311.97923 & 34.771038 & 1869489385821202816 & 0.488582 & 0.115 & 2.110629 & 0.109  \\
312.564502 & 12.412033 & 1760943570684334208 & 3.047312 & 0.209 & 15.580245 & 0.076  \\
312.603366 & 47.202077 & 2166511040319614208 & 0.38052 & 0.147 & 1.264149 & 0.127  \\
313.727051 & 39.848684 & 1872913471188045184 & 2.26941 & 0.054 & 0.678956 & 0.023  \\
314.203035 & 66.204173 & 2245787855903677952 & 1.313048 & 0.21 & 1.250967 & 0.228  \\
315.998617 & 47.984506 & 2165439879783209344 & 3.822778 & 0.095 & 2.650119 & 0.037  \\
316.499668 & 36.761395 & 1868409321799840768 & 0.543036 & 0.063 & 0.725643 & 0.036  \\
316.935639 & 47.04712 & 2165141946503725056 & 1.272084 & 0.103 & 3.086064 & 0.035  \\
317.167097 & -8.177796 & 6897094092939104768 & 0.223992 & 0.027 & 0.622656 & 0.041  \\
317.354427 & 55.485175 & 2176903937763785088 & 3.446506 & 0.256 & 3.158925 & 0.236  \\
318.20909 & 56.26039 & 2177329891137899904 & 0.90333 & 0.253 & 0.515231 & 0.087  \\
318.442107 & 51.071054 & 2166229226047556096 & 15.0598 & 0.169 & 3.180452 & 0.086  \\
\hline
\end{tabular}
\end{table*}

\begin{table*}
\centering
\ContinuedFloat
    \small
    \caption{Continued}
\begin{tabular}{ccccccc}
RA {[}J2000.0{]} & DEC {[}J2000.0{]} & Gaia dr3 source ID & $\mathrm{P_{A} [d]}$ & $\mathrm{D_{A}}$ & $\mathrm{P_{B} [d]}$ & $\mathrm{D_{B}}$ \\ 
\hline \hline
321.274386 & 49.322575 & 2170863908089217280 & 3.714414 & 0.154 & 2.643588 & 0.078  \\
322.785455 & 37.809557 & 1952168090372267008 & 1.158128 & 0.035 & 1.25026 & 0.023  \\
322.795044 & 41.61673 & 1967117802081173504 & 4.672324 & 0.102 & 3.584245 & 0.036  \\
324.064461 & 40.289056 & 1966045232784781952 & 0.264192 & 0.119 & 0.779663 & 0.061  \\
325.003123 & 56.109079 & 2178057359827382144 & 3.296832 & 0.225 & 1.291317 & 0.036  \\
325.081161 & 48.44538 & 1978029359778790016 & 0.678389 & 0.01 & 0.746095 & 0.054  \\
325.126003 & 39.737305 & 1953928099249228032 & 1.068532 & 0.22 & 1.123977 & 0.027  \\
325.523023 & 58.222165 & 2178581380197081216 & 1.810504 & 0.115 & 0.707697 & 0.062  \\
325.791627 & 27.217748 & 1800509737128120320 & 0.85369 & 0.239 & 3.215801 & 0.166  \\
326.593483 & 49.781513 & 1978958103506727424 & 0.421342 & 0.068 & 1.667499 & 0.039  \\
327.020592 & 62.78985 & 2216420454386370176 & 2.14556 & 0.228 & 0.404213 & 0.082  \\
327.271647 & 58.943552 & 2202597664775578112 & 1.06056 & 0.143 & 2.05528 & 0.107  \\
328.521806 & 57.784601 & 2199284050276697216 & 0.433052 & 0.121 & 1.459645 & 0.106  \\
329.000459 & 52.050747 & 1981103972253289088 & 0.685961 & 0.114 & 0.739756 & 0.027  \\
329.356711 & 45.001855 & 1973338392191011584 & 0.397442 & 0.112 & 6.522002 & 0.093  \\
329.904956 & 43.718061 & 1961000256819235200 & 0.473181 & 0.095 & 0.964035 & 0.103  \\
330.10459 & 57.341933 & 2199104211409698816 & 0.979846 & 0.194 & 0.443703 & 0.067  \\
330.629145 & 55.984692 & 2198180415480756096 & 1.328192 & 0.0 & 22.393498 & 0.086  \\
332.366702 & 56.166443 & 2197966251234148352 & 2.705704 & 0.168 & 2.822163 & 0.04  \\
332.720979 & 59.64723 & 2199925099913441920 & 2.15344 & 0.175 & 1.57226 & 0.067  \\
332.943294 & 54.42099 & 2005474711889658624 & 3.234134 & 0.101 & 0.617324 & 0.135  \\
333.801196 & 53.749768 & 2004590532743751424 & 0.340078 & 0.058 & 2.522543 & 0.034  \\
334.655366 & 55.538438 & 2005911802117955968 & 1.508362 & 0.197 & 0.947192 & 0.097  \\
335.103223 & 36.014911 & 1905994992912248448 & 0.834723 & 0.19 & 0.422142 & 0.116  \\
335.115083 & 24.563415 & 1878929101147587200 & 2.107929 & 0.256 & 2.068346 & 0.048  \\
337.436575 & 54.279883 & 2001849763802035840 & 2.263908 & 0.235 & 0.545817 & 0.03  \\
337.832153 & 60.857037 & 2201723866572302976 & 2.627162 & 0.107 & 101.237102 & 0.125  \\
339.088072 & 47.538959 & 1986716807300365696 & 6.451769 & 0.218 & 0.164641 & 0.043  \\
339.963254 & 54.98005 & 2003428628134264448$^{*T}$ & 2.184788 & 0.145 & 88.853882 & 0.143  \\
340.921211 & 57.261299 & 2007240188275431936 & 1.625996 & 0.064 & 1.379002 & 0.038  \\
341.302946 & 56.012542 & 2003907912126462592 & 0.404544 & 0.102 & 2.944626 & 0.058  \\
341.422596 & 53.230019 & 2002122850696945664 & 3.484958 & 0.063 & 2.227952 & 0.036  \\
342.74 & 58.714103 & 2007474143733297152 & 1.63917 & 0.116 & 1.617598 & 0.023  \\
344.803515 & 48.306854 & 1984790600366568576 & 0.699564 & 0.452 & 1.513966 & 0.034  \\
347.564145 & 65.090617 & 2208692193312331904 & 5.095412 & 0.059 & 1.198208 & 0.03  \\
349.093506 & 66.228219 & 2210495392378660864 & 2.568998 & 0.241 & 285.579588 & 0.218  \\
350.55902 & 56.248535 & 1997226175657082496 & 0.351738 & 0.091 & 1.855745 & 0.043  \\
351.050509 & 59.819433 & 2010769070836151424 & 1.312382 & 0.17 & 0.89549 & 0.057  \\
352.521122 & 57.484618 & 1998851258144250112 & 0.280512 & 0.094 & 1.619602 & 0.047  \\
352.593979 & 48.961381 & 1942119756681849216 & 4.303021 & 0.4 & 0.149352 & 0.039  \\
353.524226 & 36.870485 & 1918839728265382912 & 0.366704 & 0.208 & 4.201864 & 0.116  \\
358.069196 & 62.459358 & 2012994482376100736 & 1.950456 & 0.256 & 34.404796 & 0.119  \\
358.221987 & 61.525403 & 2012695312133566208 & 3.21884 & 0.179 & 2.533716 & 0.056  \\
\hline
\end{tabular}
\end{table*}

\end{document}